\documentclass[12pt,preprint]{aastex}
\defcitealias{paper}{KT}
\newcommand\rarm{r_{\rm arm}}%
\newcommand\comp{{\rm comp}}%
\newcommand\kmps{\mbox{km\,s${}^{-1}$}}%
\newcommand\Msun{\mbox{$M_\sun$}}%
\newcommand\Mspy{\Msun\,yr${}^{-1}$}%

\shorttitle{CHARACTERISTICS OF EVOLVED BINARY SYSTEMS}
\shortauthors{KIM \& Taam}

\begin{document}
\title{A New Method of Determining the Characteristics of Evolved 
  Binary Systems Revealed in the Observed Circumstellar Patterns: 
  Application to AFGL 3068}
\author{Hyosun Kim\altaffilmark{1} and Ronald E. Taam\altaffilmark{1,2}}
\altaffiltext{1}{Academia Sinica Institute of Astronomy and Astrophysics, 
  P.O. Box 23-141, Taipei 10617, Taiwan; hkim@asiaa.sinica.edu.tw}
\altaffiltext{2}{Department of Physics and Astronomy, Northwestern University,
  2131 Tech Drive, Evanston, IL 60208; r-taam@northwestern.edu}

\begin{abstract}
The binary characteristics of asymptotic giant branch (AGB) stars are 
imprinted in the asymmetric patterns of their circumstellar envelopes (CSEs). 
We develop a simple method for constraining the orbital parameters of 
such binary stars from the characteristics of a spiral-like pattern 
observed at large distances from the central stars. We place constraints 
on the properties of AFGL 3068 (i.e., the masses of binary components, 
the viewing inclination of the orbital plane, as well as the orbital 
period, velocity, and separation). In particular, the mass of the 
companion star of AFGL 3068 is estimated to be greater than 2.6\Msun. 
This method is applicable to other AGB stars, providing a step toward 
understanding the role binary stars can play in explaining the diverse 
patterns in observed CSEs.
\end{abstract}

\keywords{binaries: general --- 
  circumstellar matter --- 
  stars: AGB and post-AGB --- 
  stars: carbon --- 
  stars: late-type --- 
  stars: winds, outflows}

\section{INTRODUCTION}\label{sec:intro}
%------------------------------------------------------------------------------
At the end of a star's life, low- to intermediate-mass stars 
return their mass to interstellar space by means of slow, dense 
winds \citep{hab96} during the AGB phase. The material expelled 
from these evolved stars can take the form of irregular/regular 
patterns and shapes prior to mixing with the interstellar medium. 
The dense CSEs often show strong asymmetries in the late stellar
evolutionary stages, whose origin probably lies in their progenitor 
systems. Recent studies reveal that a binary system may play a 
crucial role in forming the various shapes of CSEs \citep{hug07}. 
However, the determination of the system properties of a wide 
binary is challenging. 

AFGL 3068 is the first AGB binary system that clearly shows a wide 
spiral pattern. Two point-like sources in near-infrared images have 
been revealed using the Keck adaptive optics \citep{mor06}, making 
AFGL 3068 an evolved stellar system in which the binary components 
are indeed detected. Based on a distance of 1\,kpc, the projected 
separation is 109\,AU, which is relatively large to infer a direct 
binary interaction. Nevertheless, AFGL 3068 exhibits evidence of a 
binary interaction in view of the large scale spiral pattern within 
its CSE extending over 10,000 AU in dust-scattered Galactic light 
\citep{mau06}. This spiral pattern provides support for the role of 
a companion star in shaping the CSE \citep{sok94,mas99,he07,edg08}, 
necessitating an attempt to constrain binary properties from the 
large scale density pattern \citep{kim11,kim12,paper}.

In a parameter study of hydrodynamical simulations on the asymmetric
circumstellar pattern due to the binary motions \citep[\citetalias{paper} 
hereafter]{paper}, we have shown the directional dependence of the wind
flow blown out from the AGB star. It reflects the difference between 
the line-of-sight velocity of the flow and the propagation speed of 
the pattern in the plane of the sky. The pattern propagation speed is 
$<V_w>+2V_p/3$ throughout the orbital plane, but equals to $<V_w>$ in 
the vertical direction, resulting in an overall oblate shape. This 
shape is characterized by the aspect ratio $a/b$ corresponding to 
$(<V_w>+2V_p/3)/<V_w>$, which appears as an ellipse with inclination.
In addition, the maximum fluctuation of the gas velocity relative 
to the average wind velocity value $<V_w>$ is less than the orbital 
velocity $V_p$ of the mass-losing star. 

\begin{figure*} %fig1
  \plotone{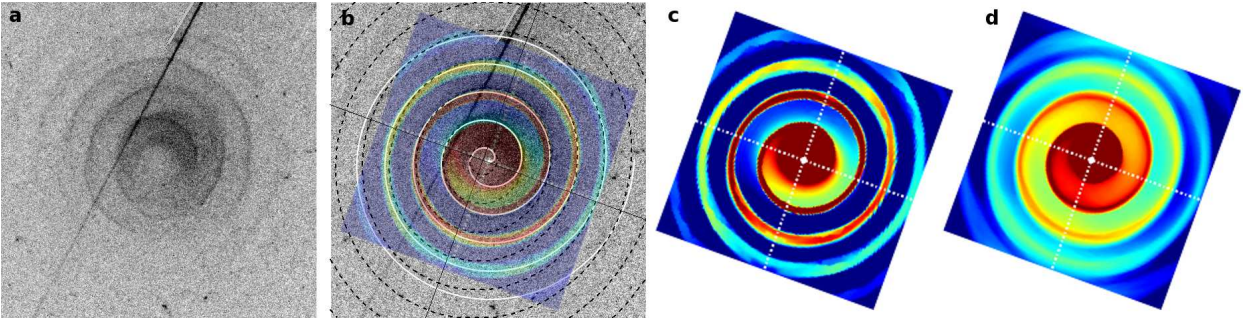}
  \caption{\label{fig:a3068}
    (a) Image of AFGL 3068 in the wide-V F606W filter of the 
    Hubble Space Telescope-ACS camera with the field size of 
    $25.\!\!\arcsec6\times25.\!\!\arcsec6$ (Credit: Mauron 
    \& Huggins, \aap, 452, 257, 2006, reproduced with permission 
    \copyright\ ESO.). North is up, and east is left. The bright 
    straight line refers to a diffraction spike of a star in the 
    field of view of the original image. 
    (b) Same image with an Archimedes spiral fit \citep{mau06}, overlaid 
    with our three-dimensional hydrodynamic model. The observed pattern 
    is more elongated than an Archimedes spiral having a constant spacing, 
    particularly in the northwest direction. The elongated shape of the 
    spiral pattern is described by the ellipses with the axial ratio of 
    1.1 corresponding to the eccentricity of 0.417 (dashed lines). The 
    ellipses have the major axis (the intersection of the plane of the
    sky; line of nodes) at 70\arcdeg\ counterclockwise from east-west, 
    where the binary stars are aligned at the center \citep{mor06}, and 
    the minor axis is perpendicular to the major axis (black thin lines). 
    The difference in the semi-major radius between the dashed lines is 
    2350\,AU at a distance of 1\,kpc.
    (c) Density distribution at the mid-plane in the line of sight of a
    hydrodynamic model for binary stars composed of a mass-losing star 
    (3.5\Msun) and a main-sequence star (3.1\Msun) with the orbital 
    separation of 166\,AU. The corresponding orbital velocities are 
    2.8\,\kmps\ and 3.2\,\kmps, respectively. The wind velocity varies 
    about a mean value of 11.5\,\kmps. An adiabatic index $\gamma$ of 1.4 
    is employed to produce a temperature profile of the form $r^{-0.83}$ 
    as obtained from the CO molecule excitation analysis \citep{woo03}. 
    The orbital plane is inclined by 50\arcdeg\ from the plane of the 
    sky.
    (d) Column density distribution.
  }
\end{figure*}

Based on these findings, we propose a new method to constrain the 
binary properties from four observed quantities corresponding to 
the projected axial ratio $(a/b)_{proj}$, the projected binary 
separation $(r_p+r_{\rm comp})_{proj}$, the position angle of the 
star from the major axis of the elliptical shape (PA), and the 
arm spacing $\Delta r_{\rm arm}$. Here, $r_p$ and $r_{\rm comp}$ 
correspond to the separation of the AGB star and its companion 
from the center of mass of the system. For the first application 
of this method, we place constraints on AFGL 3068, as an extreme 
carbon star \citep{leb77,jon78,sil79,jew82,vol92,zha09}, at the tip 
of the AGB. Figure\,\ref{fig:a3068} shows the HST image of AFGL 3068, 
overlaid with our binary model with the mass ratio of 1.13. 

% figure 1

\section{NEW METHOD}
%------------------------------------------------------------------------------
% figure 2
\begin{figure} %fig2
  \plotone{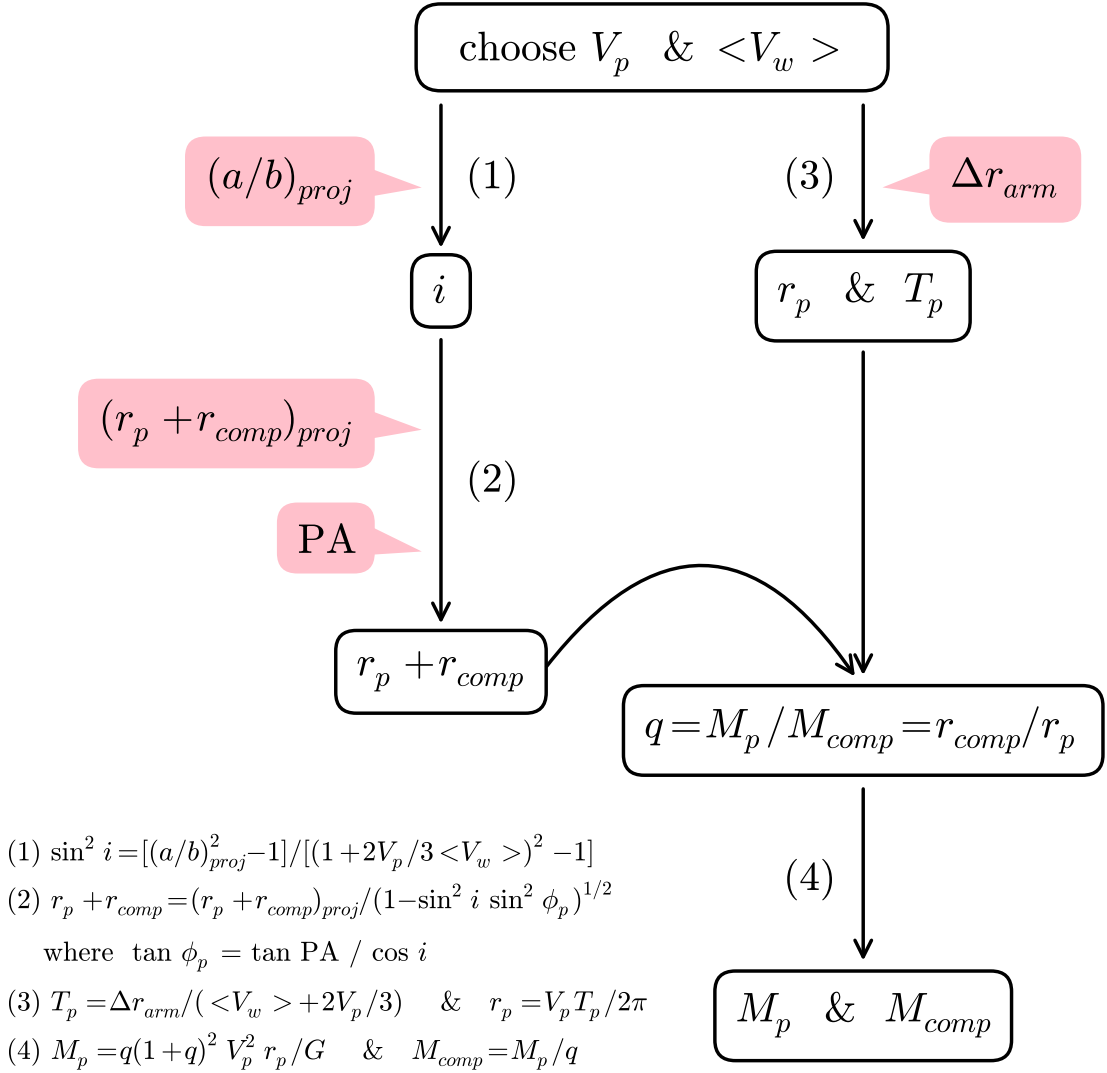}
  \caption{\label{fig:chart}
    Schematic diagram for the process of deriving binary properties
    (open boxes) from four observables (pink boxes), based on a
    parameter study using hydrodynamic simulations \citepalias{paper}.
    The speeds of orbital motion $V_p$ and wind motion $<V_w>$ of 
    the mass-losing star are considered as the model parameters. 
    The four required observables correspond to the projected 
    major-minor axial ratio of the ellipse characterizing the 
    overall shape of the pattern $(a/b)_{proj}$, the projected 
    orbital separation $(r_p+r_{\rm comp})_{proj}$, the position 
    angle of the stars with respect to the major axis (line of 
    nodes), and the spiral arm spacing $\Delta r_{\rm arm}$. 
    The result of the analysis leads to the masses $M_p$ and 
    $M_{\rm comp}$, orbital radii $r_p$ and $r_{\rm comp}$ for 
    the mass-losing star and the companion, respectively, the 
    orbital period $T_p$, and the inclination angle of the 
    orbital plane $i$.
  } 
\end{figure}

Figure\,\ref{fig:chart} is a schematic diagram summarizing the procedure
of constraining the binary stellar and orbital properties, which takes 
$V_p$ and $<V_w>$ as two free parameters. Consider an oblate spheroid 
given by the relation $(x/a)^2+(y/a)^2+(z/b)^2=1$ to describe the overall 
shape of the circumstellar pattern 
in a three-dimensional space. The determination of the projection is 
key in connecting the observed features to the binary properties.
The projected shape of the oblate spheroid is obtained by coordinate 
transformation to the coordinate rotated with the inclination angle 
$i$. Defining the $x$-axis as the line of nodes (i.e., $x_{proj}=x$ 
and $y_{proj}=y\cos i+z\sin i$) yields the elliptical shape in the 
projected plane as $(x_{proj}/a_{proj})^2+(y_{proj}/b_{proj})^2=1$ 
where $a_{proj}=a$ and $b_{proj}^{-2}=(\cos i/a)^2+(\sin i/b)^2$, 
which simplifies to
\begin{equation}\label{eqn:incli}
  \sin^2i=\frac{(a/b)_{proj}^2-1}{(a/b)^2-1}.
\end{equation}
Hence, the orbital inclination can be estimated from the observed 
(projected) axial ratio, given the relation between the aspect 
ratio of the oblate spheroid and the orbital-to-wind velocity 
ratio, $a/b=1+2V_p/(3<V_w>)$, as found from our parameter study 
in \citetalias{paper}.

% figure 3
\begin{figure} %fig3
  \epsscale{0.9}
  \plotone{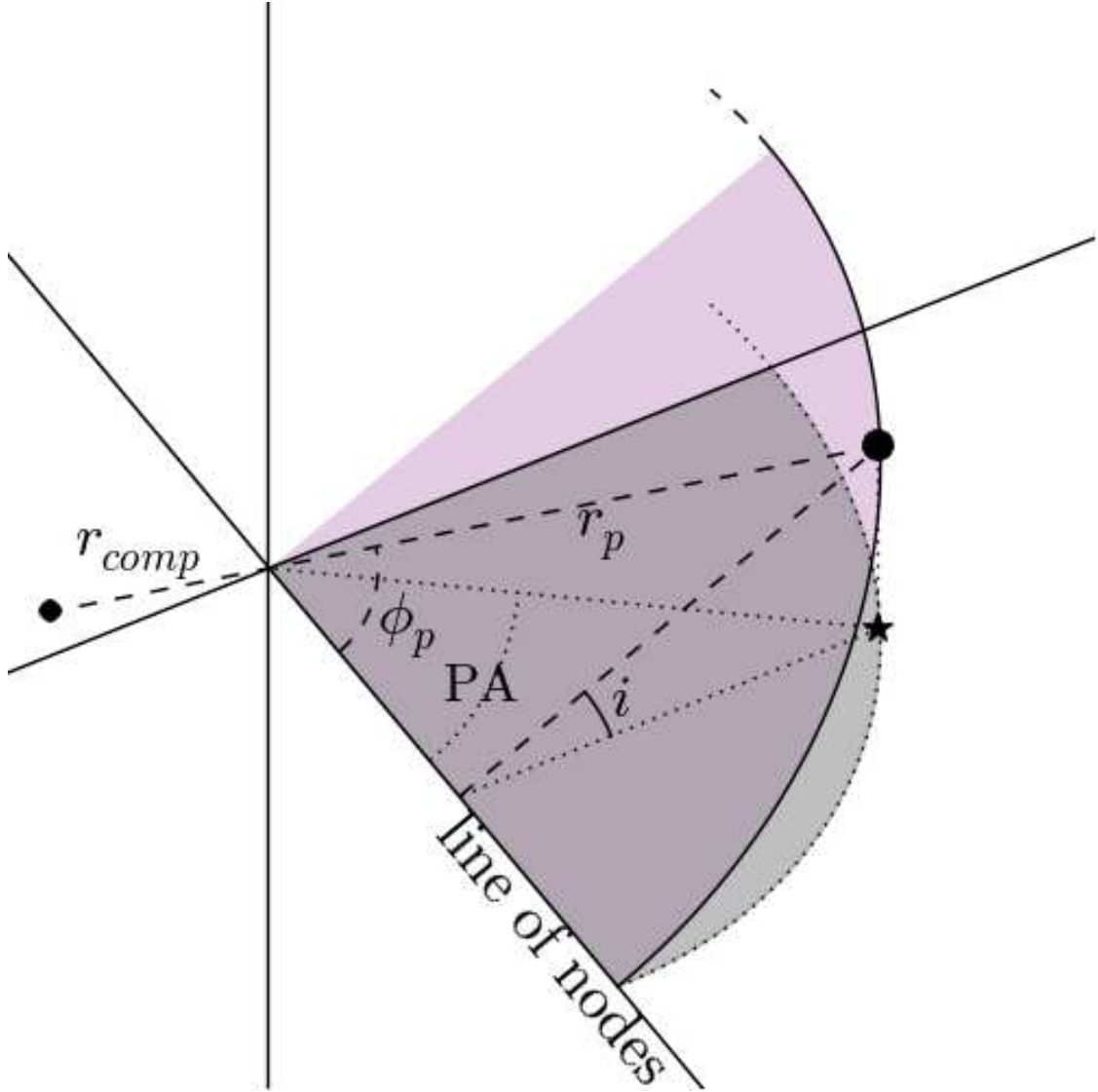}
  \caption{\label{fig:geome}
    Sketch illustrating the binary orbital plane inclined from the 
    plane of the sky with the inclination $i$. The separation of the 
    mass-losing star and companion from the common center of mass are 
    $r_p$ and $r_{\rm comp}$, respectively. The longitudinal angle 
    from the line of nodes is $\phi_p$ in the orbital plane, and the 
    projected angle in the plane of the sky is defined by PA.
  }
\end{figure}

Similarly, simple geometry yields the orbital separation of the
binary stars ($r_p+r_{\rm comp}$) as a function of the inclination 
$i$. Figure\,\ref{fig:geome} illustrates the inclined orbit with 
respect to the plane of the sky, and we obtain the projection
effect for the orbital radius of a star to be given by
$r_{p,\,proj}=r_p(1-\sin^2\phi_p\sin^2i)^{1/2}$
and the same for the companion's orbital radius $r_{\rm comp}$.
Therefore, the true orbital separation can be obtained using the 
observed separation $(r_p+r_{\rm comp})_{proj}$ and position angle 
PA of the stars from
\begin{eqnarray}\label{eqn:separ}
  \lefteqn{r_p+r_{\rm comp}=\frac{(r_p+r_{\rm comp})_{proj}}{(1-\sin^2\phi_p\sin^2i)^{1/2}}},\\
  &&\mathrm{where\ }\tan\phi_p=\tan({\rm PA})/\cos i,\nonumber
\end{eqnarray}
and the inclination derived from Equation (\ref{eqn:incli}).

On the other hand, the arm spacing $\Delta\rarm$ of the spiral pattern 
in a binary system is determined by the product of the binary orbital 
period and the pattern propagation speed in the orbital plane:
\begin{equation}\label{eqn:drarm}
  \Delta\rarm=\left(<V_w>+\frac{2}{3}V_p\right)\times\frac{2\pi r_p}{V_p}.
\end{equation}
This equation shows that the orbital radius $r_p$ can be obtained 
from the pattern spacing $\Delta\rarm$ along the observed longest axis, 
which refers to a section of the orbital plane, for a given velocity 
ratio $V_p/<V_w>$. Using $r_p+r_{\rm comp}$ and $r_p$ from Equations 
(\ref{eqn:separ}) and (\ref{eqn:drarm}), the binary mass ratio 
$q=M_p/M_{\rm comp}$ corresponding to $r_{\rm comp}/r_p$ is determined.

Finally, we employ the formula for the force balance in a binary system,
\begin{equation}
  \frac{V_p^2}{r_p}=\frac{GM_{\rm comp}}{(r_p+r_{\rm comp})^2}
\end{equation}
and replace the quantities related to the companion by the quantities of 
the mass-losing star using the definition of the $q$ factor, resulting in 
$M_p=q(1+q)^2V_p^2r_p/G$. The companion mass is $M_p/q$.

\section{CASE STUDY OF AFGL 3068}
\subsection{Observed Properties}
%------------------------------------------------------------------------------
We first find the best fit of an elliptical shape with the dust 
scattered light pattern of AFGL 3068 (Figure\,\ref{fig:a3068}(b)). 
The modeled ellipse is characterized by an axial ratio $(a/b)_{proj}$ 
of 1.1 with the major axis in 70\arcdeg\ from the west (i.e., right). 
Identifying the major axis as the line of nodes of the inclined orbital 
plane for the reference direction, the position angle (PA) of two 
point-like sources arranged to the east-west direction is 70\arcdeg. 
The observed separation between two sources $(r_p+r_{\rm comp})_{proj}$ 
is reported \citep{mor06} to be 109\,AU\,($d$/1\,kpc), and the interval 
between ellipses along the major axis, $\Delta r_{\rm arm}$, is 
2350\,AU\,($d$/1\,kpc), where $d$ refers to the object distance. 
The slight deviation from a very good match of the elliptical shape 
with most parts of the spiral pattern implies that partial circular 
(or elliptical) rings observed in other AGB CSEs could also contribute 
to the spiral pattern driven by a binary motion. 

\subsection{Parameter Space}
%------------------------------------------------------------------------------
% figure 4
\begin{figure} %fig4
  \plotone{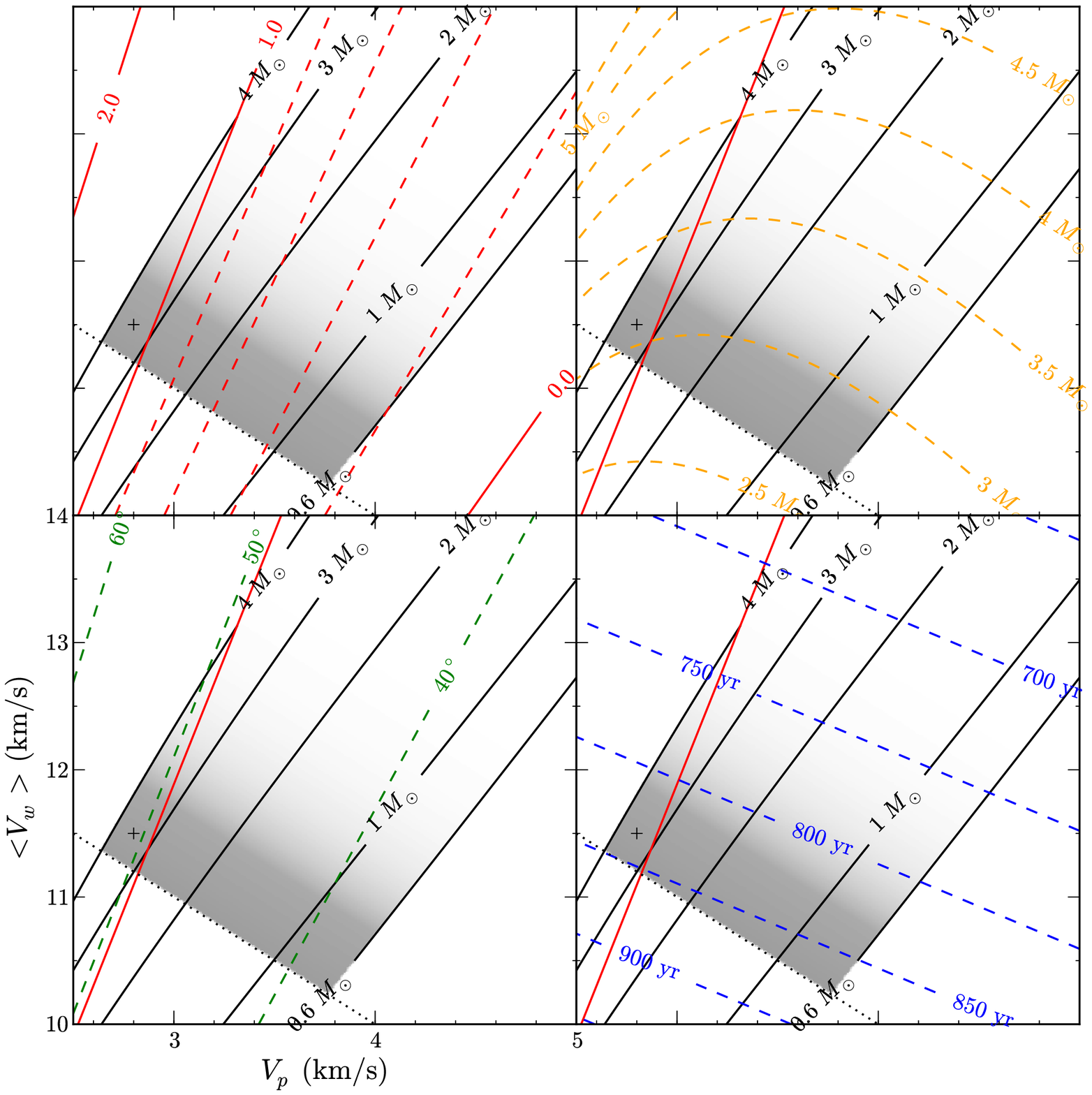}
  \caption{\label{fig:degen}
    Model parameter space for the speeds of orbital motion $V_p$ and
    outflowing wind $<V_w>$ of the mass-losing star in AFGL 3068. An 
    expansion velocity obtained from molecular line emissions \citep%
    {lou93} provides a constraint, $<V_w>+V_p\gtrsim14$\,\kmps\ (dotted 
    line), and the mass of the AGB carbon star \citep{boo93,boo95} 
    can be 0.6--4\Msun\ (black solid line), confining the parameters
    for AFGL 3068 into the gray-colored region. Considering the carbon 
    star to be more massive than its companion, the parameter space 
    for AFGL 3068 is further restricted to above the red line for the 
    mass ratio of 1. The $+$ sign corresponds to the parameters for 
    which a hydrodynamic simulation was carried out for comparison to 
    observations (Figure\,\ref{fig:a3068}).
    \textbf{Top left.} 
    The mass ratio between the carbon star and companion of 0, 1, and 2 
    (red solid lines) and 0.2, 0.4, 0.6, and 0.8 from bottom (red dashed 
    lines). 
    \textbf{Top right.} 
    The mass of the companion with an 0.5\Msun-interval (yellow dashed lines).
    \textbf{Bottom left.}
    The inclination angle of the orbital plane with respect to the 
    plane of the sky with a ten-degree interval (green dashed lines).
    \textbf{Bottom right.}
    The orbital period of binary stars with a 50-year interval 
    (blue dashed lines).
  }
\end{figure}

For AFGL 3068, we construct a parameter space, in which $V_p$ and $<V_w>$ 
are spanned, and follow the procedure summarized in Figure\,\ref{fig:chart}
to calculate the corresponding orbital properties of the binary system. 
Figure\,\ref{fig:degen} shows the result, restricting the parameter space
that satisfies the conditions for AFGL 3068 as a carbon-rich AGB star in 
a binary system. First, the gas velocity cannot exceed $<V_w>+V_p$ in any 
direction \citepalias{paper}, which yields $<V_w>+V_p\ga14$\,\kmps\ with the 
line-of-sight velocity of 14\,\kmps\ defined from the half width of spectral 
lines \citep{lou93}. Second, the mass of a carbon star such as AFGL 3068 
is theoretically constrained to be less than 4\Msun\ since more massive 
stars (4 to 8\Msun) are sufficiently hot at the base of their convective 
envelopes to activate the so-called ``hot bottom burning'' process, which 
significantly burns the envelope carbon \citep{boo93,boo95}. Third, we 
take the lower limit of the carbon star as 0.6\Msun, which is the mean 
mass of white dwarfs \citep{mad04}. 

These three constraints place the 
AFGL 3068 system within the gray-colored area in Figure\,\ref{fig:degen}. 
As a result, the mass of the companion is constrained to be greater than 
2.6\Msun\ with the binary mass ratio $q=M_p/M_{\rm comp}$ greater than 
0.2. In addition, the inclination angle of the system is constrained 
to lie within approximately 40\arcdeg--50\arcdeg. Finally, the orbital 
period is found to be less than 870 years, and the orbital speed of the 
carbon star in the range of 2.5--4\,\kmps. Assuming an accuracy of 
the distance measurement to be 20\%, the companion mass is estimated to 
be greater than 2.1 to 3.2\Msun. If the line-of-sight velocity of the 
wind is 16\,\kmps\ \citep{deb10}, the estimated companion mass is larger 
than 2.8 or 4.2\Msun\ at distance of 0.8 or 1.2\,kpc. The combination 
of a high wind velocity (16\,\kmps) and large distance (1.2\,kpc) results 
in a large companion mass, leading to an implausible mass $>$\,4\Msun, 
for which the companion should have evolved more rapidly than the carbon 
star.

As more massive stars evolve more rapidly, the carbon star is likely to 
be more massive than its companion ($q>1$), although the mass ratio $q$ 
could be less than unity because the carbon star is in the phase of losing 
mass at a high rate ($>10^{-5}$\,\Mspy) \citep{jon78,vol92,woo03,lou93,
ram08,lad10}. In the case that the total mass lost from the carbon star is 
not too extensive, the parameters can be further constrained into a small 
parameter space, indicating individual stellar masses of 3--4\Msun, speed 
of 2.5--3.5\,\kmps\ with the orbital period of 700--850 years in the plane 
inclined by approximately 50\arcdeg.

\subsection{Hydrodynamic Model}
%------------------------------------------------------------------------------
Within the limited area in the parameter space where AFGL 3068 
can lie, we select a point (plus sign in Figure\,\ref{fig:degen}) 
to compare the corresponding hydrodynamic model with the observed 
pattern in the CSE of AFGL 3068.
Figure\,\ref{fig:a3068} indeed shows that the model shows good agreement 
with the scattered light image of AFGL 3068. In particular, the Archimedes 
spiral characterized by having a constant spacing \citep{mau06} passes 
inside the intensity peak positions in the northerly direction, while 
the spiral of our model better traces the positions of the pattern. 
The blurred patterns seen in the northwest and south closely resemble 
the knotty structures appearing along the line of nodes, which form from 
the overlap of the two structures generated by the orbital motions of 
individual stars \citepalias{paper}. 

\begin{figure} %fig5
  \plotone{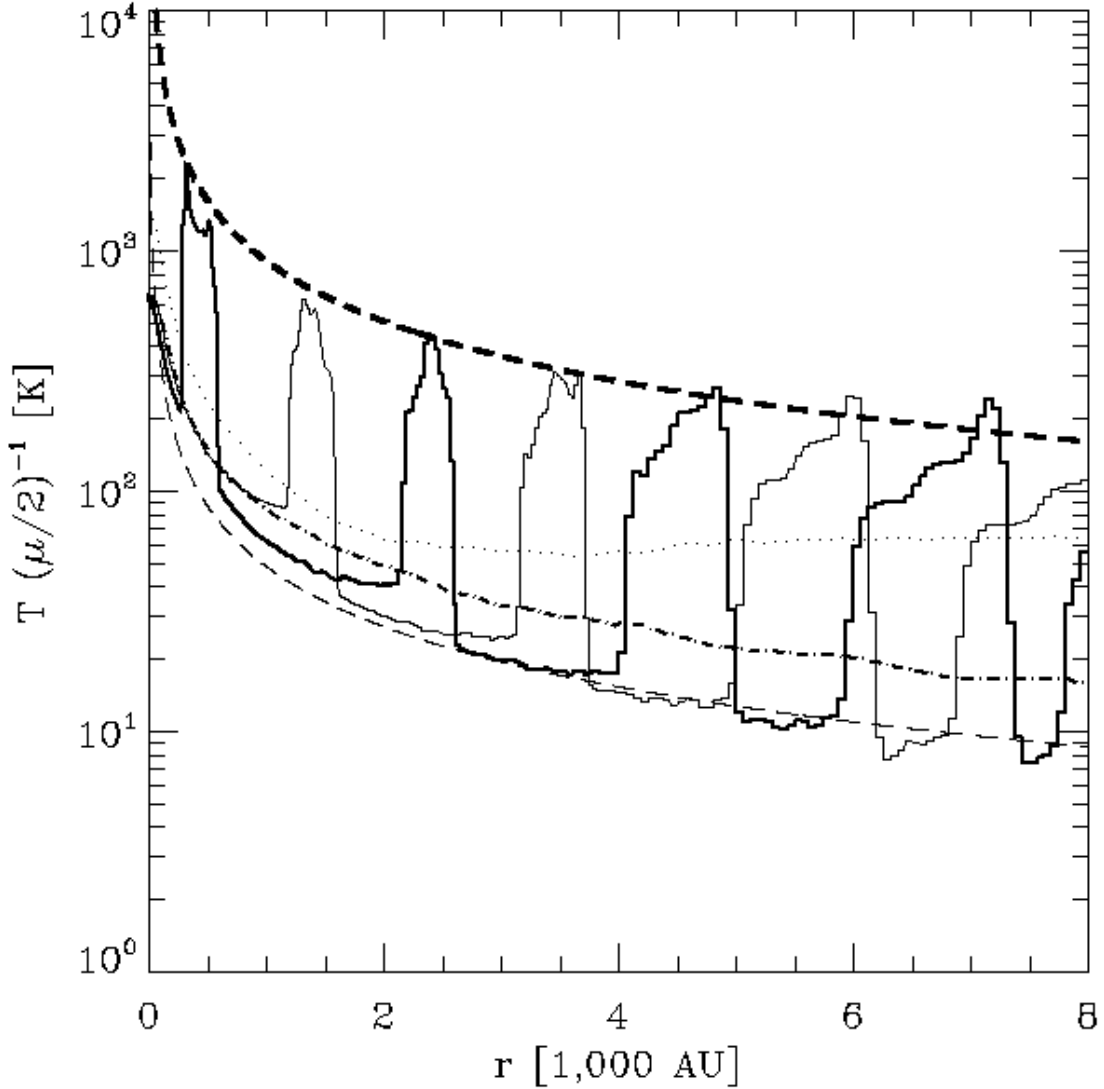}
  \caption{\label{fig:tempe}
    Temperature profile in the outflowing CSE 
    surrounding the binary system along two arbitrary-chosen 
    directions in the orbital plane (thick and thin solid lines) in the 
    hydrodynamical model. The peak temperature within the spiral arm 
    has the value $2.8\times10^5$\,K\,$(\mu/2)(r/{\rm AU})^{-0.83}$ 
    (thick dashed line), while the inter-arm temperature approximates 
    to $1.5\times10^4$\,K\,$(\mu/2)(r/{\rm AU})^{-0.83}$ (thin 
    dashed line). The temperature along the orbital axis (dot-dashed 
    line) and the value averaged over three dimensional space (dotted 
    line) are consistent with the result of a radiative transfer modeling 
    for the observed CO line emission \citep{woo03}.
  }
\end{figure}

The modeled binary system is characterized by a mass-losing 
star of mass $M_p=3.5\Msun$, velocity $V_p=2.8\,\kmps$, and orbital 
radius $r_p=78$\,AU, and a companion of mass $M_\comp=3.1\Msun$, 
velocity $V_\comp=3.2\,\kmps$, and orbital radius $r_\comp=88$\,AU. 
The wind velocity $V_w$ was set to be 10.9\,\kmps\ at 30\,AU distance 
from the mass-losing star. The gas flow evolves to a quasi-steady state 
fluctuating about 11.5\,\kmps\ with a standard deviation of 0.67\,\kmps.
To achieve a temperature profile of $r^{-0.83}$ as determined from an 
analysis of the CO molecule excitation \citep{woo03} for a density 
profile of $r^{-2}$, the adiabatic index $\gamma$ of the wind material 
is set to be 1.4 with the initial temperature of 1540\,K\,$(\mu/2)$ at 
30\,AU distance from the mass-losing star, where $\mu$ indicates the 
molecular weight. Figure\,\ref{fig:tempe} shows that the resulting 
temperature profile indeed follows the expected power law with the index 
of $-0.83$ and has an average value of approximately 60\,K\,$(\mu/2)$. 
Additional heating in the post-shock region and cooling via molecular 
line transitions are not considered as the timescales in the CSEs 
of AGB stars are comparable \citep{edg08}.
The density power law index is $-2.2$ in the resulting configuration. 
The volume and column density contrasts correspond to a factor of 10--30 
and 2--5, respectively. We note that the density contrast in the model 
with $\gamma=1.4$ is approximately constant, while the contrast in an 
isothermal model decreases rapidly with distance from the system center 
\citepalias{paper}. The dependencies of the overall shape and morphology of 
the spiral pattern on the orbital-to-wind velocity ratio are insensitive to 
the value of $\gamma$. The mass loss rate taken as $4\times10^{-5}$\,\Mspy\ 
\citep{oli01} provides the scale for the magnitude of the density, but not 
the density contrast. The resulting column density toward the star is of 
order of $10^{23}$\,cm$^{-2}$, which varies by a factor of a few with the 
inclination angle of the system.

% figure 5

The observed pattern thickness of AFGL 3068 is thinner than the arm
thickness of the modeled column density distribution, and is rather
comparable to the density distribution in the mid-plane of the sky. 
Because of the curved geometry of the spiral-shell pattern, the shocked
edges of the pattern at different layers are projected into different
positions in the plane of the sky, resulting in a broader arm for the 
column density. It may imply that the observed dust scattered light is 
optically thick so that the observed image refers to only part of the 
pattern in the line of sight. This is consistent with the high optical
depth derived from the spectral energy distribution fitting \citep{lad10}.
The observed pattern thickness may also depend on the observed waveband.
In particular, the size distribution of the dust grains may be affected, 
resulting from coagulation or shattering via grain-grain collisions 
immediately after encountering the shock with a velocity dispersion of 
a few \kmps\ \citep{jon96}. As a consequence, the variation of the size 
distribution may affect the intensity across the shock pattern at the 
observed wavelength. Small grains with a size less than 0.045\,\micron\ 
tend to strongly couple with the gas, while larger grains are characterized 
by a different distribution as they are more resistant to the drag force 
\citep{van11}. The probable large population of small grains near the shock 
increases the scattering cross section with the large surface-to-volume 
ratio of the small grains, resulting in an increase of the scattered light 
intensity at the corresponding wavelengths.

Our model assumes circular motion of the binary stars. However, previous 
calculations for eccentric orbits do not show a noticeable change in 
the overall shape outside the second turn of the spiral pattern \citep%
{he07,rag11}, suggesting that our model for the large scale shape of the 
pattern driven by circular orbits is a good approximation even for the 
case of non-circular orbits. Although a highly eccentric orbit ($e\ga0.8$) 
causing substructure in the inter-arm region (prominent between the 
innermost turns) may be responsible for the extra 
structure found in the observed image between first and second turns in 
the northwest and southeast, a circular orbit model also exhibits a weak 
density enhancement close to the observed substructure.
This substructure can also possibly be affected by local phenomena 
around the companion (e.g., the formation of an accretion disk and 
radiation pressure effects).

\section{Conclusion}
We have developed a new method to estimate the stellar and 
orbital characteristics of binary systems with an AGB component 
using information gleaned from the properties of a spiral pattern 
in the CSE. As a first application, we have 
provided constraints on the inclination angle, orbital period, 
and companion mass in the binary system AFGL 3068 based on the 
pattern shape of AFGL 3068 seen in dust scattered light. With 
molecular line observations, additional kinematic information 
on the velocity structure of the pattern will provide further 
constraints on the system properties.

The potential for estimating system parameters with the method 
proposed in this paper is promising and can be useful in providing 
fundamental input about systems where asymmetries in CSEs are 
observed. As such one may be able to use these observational data to 
place constraints on the input parameters and prescriptions used for 
the evolution of systems in population synthesis studies of detached 
binary systems \citep[see][]{pol11}. It is likely that the application 
of this method to other systems, which we can anticipate will be found 
with very sensitive and high resolution observations using e.g., the 
Atacama Large Millimeter/submillimeter Array, will provide a greater 
statistical basis for determining the incidence of such evolved binary 
systems in the total stellar binary population.

\acknowledgments
%------------------------------------------------------------------------------
This research is supported by the TIARA in the ASIAA. We are grateful 
to \aap, N.~Mauron and P.~J.~Huggins for their agreement to allow us to 
reprint their published HST image. H.K. thanks H.~Hirashita, P.~Koch, 
I-T.~Hsieh, members of ICSM and CompAS groups in ASIAA for discussions 
and comments.

\end{document}